\documentclass[twocolumn,floatfix,showpacs]{revtex4}
\usepackage{graphicx}
\usepackage{amsmath,amssymb}
\usepackage{bm}

\begin{document}
\newcommand{\be}{\begin{equation}}
\newcommand{\en}{\end{equation}}
\newcommand{\epl}{Europhys. Lett. }
\newcommand{\br}{\boldsymbol{r}}
\renewcommand{\Im}{{\rm Im}}
\renewcommand{\Re}{{\rm Re}}

\title{Entanglement entropy in Fermi gases and Anderson's orthogonality catastrophe}
\author{A. Ossipov}
\email{alexander.ossipov@nottingham.ac.uk}
\affiliation{School of Mathematical Sciences, University of Nottingham, Nottingham NG7 2RD, United Kingdom} 
\date{\today}
\begin{abstract}
We study the ground-state entanglement entropy  of a finite subsystem of size $L$ of an infinite system of noninteracting fermions  scattered by a potential of  finite range $a$. We derive a general relation between the scattering matrix and the overlap matrix and use it to prove that for a one-dimensional symmetric potential the von Neumann entropy,  the R\'enyi entropies, and the full counting statistics are robust against potential scattering,  provided that $L/a\gg 1$. The results of numerical calculations support the validity of this conclusion for a generic potential.
\end{abstract}
\maketitle

\section{Introduction}
\label{intro}
The ground state of free fermions is not robust against a perturbation by an external potential. As shown in the seminal Letter by Anderson \cite{And67},  the overlap between the ground states of  a system of noninteracting fermions in a box of size $L$ in the presence of a finite range potential $|\Psi_V\rangle$ and without a potential  $|\Psi_0\rangle$ vanishes in the limit $k_FL\to\infty$,
\be
\langle \Psi_V |\Psi_0\rangle\sim(k_FL)^{-\eta},
\en 
where  $k_F$ is the Fermi wave vector and $\eta$ can be expressed in terms of the scattering phase shifts created by the potential. This result implies that the new ground state is orthogonal to the initial one, but it does not provide any information on how the internal properties of the ground states, such as quantum correlations, are affected by an external potential.  

We address here this question investigating the bipartite entanglement of the ground states. Entanglement has recently come to be viewed as a unique feature of quantum mechanics which reveals connections between concepts of quantum information, quantum field theory, and many-body physics. In the context of condensed matter physics, entanglement entropies have been studied intensively as a measure of quantum correlations in many-body systems \cite{AFOV08}. In particular, a lot of works have been devoted to investigation of the ground-state entanglement entropy between a finite subsystem of size $L$ and the rest of an infinite system \cite{LR09, ECP10}. Such bipartite entanglement entropies can be calculated from the spectrum of the reduced density matrix of a subsystem, which in turn can be constructed from the ground state wave function.

For noninteracting fermions the ground state wave function is given by the Slater determinant made up of the single particle wave functions. Thus the knowledge of the single particle states is in principle sufficient for constructing of the reduced density matrix and, hence, for computing of the entanglement entropy. More specifically, one can first introduce the overlap matrix $A_{nm}$ built up from the single particle wave functions $\psi_n(x)$ corresponding to the lowest $N$ energy levels of an infinite system \cite{K06,CMV11}
\be
A_{nm}=\int_{-L/2}^{L/2}dx\; \psi^{\star}_{n}(x)\psi_{m}(x), \quad n,m=1,\dots, N.
\en
The finite range of integration here is due to the projection of the wave functions onto a finite subsystem, for which the entanglement entropy is calculated. 
Then the entanglement entropy $S$ of the ground state, in which the lowest $N$ energy levels are filled, can be expressed in terms of the eigenvalues $\lambda_{i}$ of $A_{nm}$
\be\label{entropy}
S=-\sum_{i=1}^N(\lambda_i\ln\lambda_i+(1-\lambda_i)\ln(1-\lambda_i)).
\en

In practice, the problem of analytical calculation of the entanglement entropy is nontrivial even in the simplest case of one-dimensional free fermions in the absence of any scattering potential. In this case, the asymptotic scaling behavior of the von Neumann entanglement entropy in the limit $k_FL\to\infty$ is given by \cite{VLRK03,JK04,GK06,CMV11,SI12}
\be\label{free-fermions}
S=\frac{1}{3}\ln(2k_FL)+\Upsilon+o(1),
\en
where the constant $\Upsilon$ is defined by Eq.(\ref{entropy_final}). The appearance of the logarithmic dependence in this expression is due to the sharpness of the Fermi surface. Mathematically, the derivation of Eq.(\ref{free-fermions}) is based either on asymptotics of the determinants of the Toeplitz matrix $A_{nm}$ for one-dimensional fermions \cite{JK04,KM05,SI12} or on the Widom conjecture for its higher dimensional generalizations  \cite{GK06,LSS13}. The translational invariance of the free fermion system is essential for application of both methods.

In this work we study the entanglement entropies of noninteracting fermions in the presence of a finite range scattering potential.  The presence of a potential breaks the translational invariance of the system and therefore makes it impossible to apply straightforwardly the techniques used in the derivation of Eq.(\ref{free-fermions}). 

Moreover the single particle wave functions are not known explicitly for an arbitrary potential and therefore the overlap matrix $A_{nm}$ cannot be constructed directly. In order to overcome this problem, we derive first a general relation between $A_{nm}$ and the scattering states, showing that the scattering matrix contains all information about the entanglement of noninteracting fermions. 

We focus then on the  case of a one-dimensional symmetric potential of finite range $a$, for which the overlap matrix takes the form of a generalized sine kernel defined in terms of the scattering phase shifts. Using the asymptotic result for the determinant of such a kernel, we are able to prove that the entanglement entropy in the limit $k_FL\to\infty$ has the same asymptotic form as for free fermions Eq.(\ref{free-fermions}).  This conclusion is remarkable  taking into account the fact that the nonzero phase shifts are responsible for orthogonality of the modified ground state \cite{And67}, they also enter explicitly into the expression for the determinant of the overlap matrix, but at the same time they do not alter its spectral density.  Furthermore, we show that not only the von Neumann entropy, but also the R\'enyi entropies and the full counting statistics are not affected by a finite range scattering potential.  The robustness of the entanglement is an important result, in particular, in view of the potential use of entanglement as a fundamental resource for quantum information.

The outline of this Letter is as follows. In Sec.~\ref{scattering}, a general relation between the overlap matrix and the scattering matrix is established for an arbitrary scattering potential. The relation is then used in Sec.~\ref{symmetric} in the case of a one-dimensional symmetric potential, in order to calculate analytically the spectral density of the overlap matrix, the entanglement, and the R\'enyi entropies, and the full counting statistics. In Sec.~\ref{disordered}, the entanglement entropy is computed numerically for a finite range disordered potential  and its scaling is compared with the results for  free fermions. Finally, our main conclusions are summarized in Sec.~\ref{conclusions}. 

\section{The overlap matrix and the scattering states}
\label{scattering}
We consider two solutions $\psi_k(\br)$ and  $\psi_{k'}(\br)$ of the $d$-dimensional Schr\"odinger equation corresponding to two different energies $\hbar^2 k^2/2m$ and $\hbar^2 k'^2/2m$
\begin{eqnarray}
\left(-\Delta+\frac{2 m}{\hbar^2}V(\br)\right)\psi_k(\br)&=&k^2\psi_k(\br),\nonumber\\
\left(-\Delta+\frac{2 m}{\hbar^2}V(\br)\right)\psi_{k'}(\br)&=&k'^2\psi_{k'}(\br),
\end{eqnarray}
where $\Delta$ is the $d$-dimensional Laplace operator and $V(\br)$ is an arbitrary potential of finite range $a$. In order to calculate the corresponding matrix element $A_{kk'}$ of the overlap matrix, we multiply  the complex conjugate  of the first equation by  $\psi_{k'}(\br)$ and the second equation by  $\psi_k^{\star}(\br)$. Subtracting one resulting equation from the other and integrating we obtain
\be
\int\limits_{B_L}d\br\left(\psi_{k'}\Delta \psi_k^{\star} -\psi_k^{\star} \Delta \psi_{k'} \right)=(k'^2-k^2)\int\limits_{B_L}d\br\,\psi_k^{\star} \psi_{k'}.
\en
The domain of integration  $B_L$ corresponds to a finite subsystem, which we are interested in, and can be chosen for example as a ball of radius $L/2$. Applying the Green's second identity to the left-hand side, we derive the following expression for the overlap matrix $A_{kk'}=\int_{B_L}d\br\,\psi_k^{\star} \psi_{k'}$:
\be\label{A_general}
A_{kk'}=\frac{1}{k'^2-k^2}\int\limits_{S_L}d\boldsymbol{\Omega}\left(\psi_{k'}\frac{\partial\psi_k^{\star}}{\partial r}-\psi_k^{\star}\frac{\partial\psi_{k'}}{\partial r}\right),
\en
where $S_L$ is a sphere of radius $L/2$ and the derivatives on the right-hand side are in the direction of the outward  normal to the sphere.  This formula shows that the knowledge of the wave functions and their derivatives at the boundary of a subsystem is sufficient for the constructing of the overlap matrix. In particular, for $L>a$ the right-hand side of Eq.(\ref{A_general}) can be expressed in terms of the scattering states and thus it provides a relation between the overlap matrix and the $S$ matrix.

To be more specific we focus now on a one-dimensional scattering potential centered at the origin, which does not support any bound states.  Each energy level $E=\hbar^2 k^2/2m$ is then twofold degenerate and there is a freedom of choosing two orthonormal states $\psi_{k}^1$ and $\psi_{k}^2$ for construction of the overlap matrix. The most convenient way to do it is to use the eigenvectors 
$f_{k}^{\alpha}=(a_{k}^{\alpha}, b_{k}^{\alpha})^{T}$ and the eigenvalues $e^{2i\delta_{k}^{\alpha}}$ of the  $S$ matrix:
\be\label{scatt_state}
\psi_{k}^{\alpha}(x)=\left\{\begin{aligned}
&b_{k}^{\alpha}\cos(kx-\delta_{k}^{\alpha}),\; x<-a/2\\
&a_{k}^{\alpha}\cos(kx+\delta_{k}^{\alpha}),\; x>a/2\\
\end{aligned}\right.\; \alpha=1,2.
\en
The overlap matrix $A_{kk'}^{\alpha \beta}$ acquires a $2\times 2$ block structure and its matrix elements can be calculated using Eq.(\ref{A_general})
\begin{eqnarray}\label{A-S_relation}
A_{kk'}^{\alpha \beta}&=&(f_{k}^{\alpha}, f_{k'}^{\beta})\left(M_{kk'}^{\alpha \beta-}+M_{kk'}^{\alpha \beta+}\right),\; 0\le k\le k_F,\nonumber\\
M_{kk'}^{\alpha \beta\pm}&=&\frac{\sin\left((k\pm k')L/2+\delta_{k}^{\alpha}\pm\delta_{k'}^{\beta}\right)}{\pi(k\pm k')}.
\end{eqnarray} 
The matrix $M_{kk'}^{\alpha \beta -}$ has the form of a generalized sine kernel \cite{KKMST09} and it reduces to the standard sine kernel, if the scattering phases 
$\delta_{k}^{\alpha}=0$. In that case $M_{kk'}^{\alpha \beta -}$ is a Toeplitz matrix and  $M_{kk'}^{\alpha \beta +}$ is its Hankel counterpart. Then the standard techniques can be used to calculate the density of the eigenvalues of $A$ asymptotically as $k_FL\to \infty$ and the free fermion result (\ref{free-fermions}) can be reproduced.

Equation (\ref{A-S_relation}) provides a general relation between the overlap matrix and the scattering matrix for an arbitrary scattering potential, allowing us to compute the entanglement entropy if the eigenvalues and the eigenvectors of the $S$ matrix are known explicitly. This relation can be easily generalized to higher dimensions. If one is interested in the scaling of the entanglement entropy without specifying the details of the $S$ matrix, then one needs to use asymptotic results for a kernel given by a sum of a generalized sine kernel and its Hankel counterpart. We are not aware of such results in a generic case, but the results for  a generalized sine kernel alone do exist in the literature \cite{KKMST09,S10,K11} and are employed in the next section. 

\section{Symmetric potential}
\label{symmetric}

For a one-dimensional symmetric potential $V(x)=V(-x)$ the eigenfunctions of the Hamiltonian are also the eigenfunctions of the parity operator $(\hat{P}\psi)(x)=\psi(-x)$ and therefore they are odd and even functions of $x$. As a result, the eigenvectors of the $S$ matrix become $k$ independent and are given by $f^{1}=(1/\sqrt{2},1/\sqrt{2} )^{T}$ and $f^{2}=(1/\sqrt{2},-1/\sqrt{2} )^{T}$. The overlap matrix in this case takes a block diagonal form with only two nonzero blocks $A_{kk'}^{\alpha \alpha}=M_{kk'}^{\alpha \alpha-}+M_{kk'}^{\alpha \alpha+}$, which correspond to even eigenstates for $\alpha=1$ and odd eigenstates for $\alpha=2$. In the absence of scattering, the corresponding phases $\delta_k^1=0$ and $\delta_k^2=\pi/2$, as it follows from Eq.(\ref{scatt_state}). It is more convenient to express them in terms of the scattering phase shifts defined as $\delta_k^+=\delta_k^1$ and $\delta_k^-=\delta_k^2-\pi/2$.

The entanglement entropy, introduced in Eq.(\ref{entropy}), can be calculated as 
\be\label{entropy_dos}
S=\int_0^1d\lambda\: \rho(\lambda) s(\lambda),\;\rho(\lambda)=\sum_{n}\delta(\lambda-\lambda_n),
\en
where $s(\lambda)=-\lambda\ln\lambda-(1-\lambda)\ln(1-\lambda)$ and $\rho(\lambda)$ is the spectral density of the overlap matrix. 
As shown in detail in the Supplemental Material \cite{suppl} the spectral density can be written as
\be\label{rho_sum}
\rho(\lambda)=\frac{1}{2}(\rho^{+}(\lambda)+\rho^{-}(\lambda)),
\en
where $\rho^{\pm}(\lambda)$ is the spectral density of the integral operators $\hat{M}^{\pm}$ with the kernels
\be
M^{\pm}(k,k')=\frac{\sin\left((k-k')L/2+\delta_{k}^{\pm}-\delta_{k'}^{\pm}\right)}{\pi(k-k')}.
\en
As shown further in \cite{suppl}  the spectral density of  $\hat{M}^{\pm}$ can be calculated using the determinants $\det\left[1+\gamma\hat{M}^{\pm}\right]$ with some parameter $\gamma$. The expression for $\rho(\lambda)$ in the limit $k_FL\to \infty$ and $k_Fa={\rm const}$ is then given  by Eq.(S12) in  \cite{suppl}:
\be
\rho(\lambda)=\frac{\ln( 2k_FL) - \Re\:\psi\left(\frac{1}{2}+\frac{i}{2\pi}\ln\frac{1-\lambda}{\lambda}\right)}{\pi^2\lambda(1-\lambda)}+o(1),
\en 
where $\psi(z)$  denotes the digamma function. Although the scattering phase shifts enter into the expression for $\det\left[1+\gamma\hat{M}^{\pm}\right]$, they do not affect $\rho(\lambda)$ and therefore this formula coincides exactly with the corresponding  result for free fermions in the absence of scattering \cite{SI12,comment}.

Substituting $\rho(\lambda)$ into Eq.(\ref{entropy_dos}), we reproduce the result for the entanglement entropy of free fermions:
\begin{eqnarray}\label{entropy_final}
S&=&\frac{1}{3}\ln(2k_FL)+\Upsilon+o(1),\nonumber\\
\Upsilon&=&\frac{2}{\pi^2}\int_0^1d\lambda \frac{\ln\lambda}{1-\lambda}g(\lambda)\approx0.495018,
\end{eqnarray}
where $g(\lambda)=\Re\:\psi\left(\frac{1}{2}+\frac{i}{2\pi}\ln\left(\frac{1-\lambda}{\lambda}\right)\right)$. This shows that Eq.(\ref{free-fermions}) is universal and not affected by the Anderson orthogonality catastrophe. 

The knowledge of the spectral density allows us also to calculate  the R\'enyi entropies:
\begin{eqnarray}
S_{\alpha}&=&\frac{1}{1-\alpha}\int_0^1d\lambda\:\rho(\lambda)\ln\left(\lambda^{\alpha}+(1-\lambda)^{\alpha}\right)\nonumber\\
&=&\frac{1+\alpha}{6\alpha}\ln(2k_FL)+\Upsilon_{\alpha}+o(1),\nonumber\\
\Upsilon_{\alpha}&=&\int_0^1d\lambda\frac{\ln\left(\lambda^{\alpha}+(1-\lambda)^{\alpha}\right)g(\lambda)}{\pi^2(1-\alpha)\lambda(1-\lambda)}.
\end{eqnarray} 
Comparing these results with ones available in the literature for free fermions    \cite{JK04,CMV11,SI12}, we reveal again the same degree of the universality of  the R\'enyi entropies.

Furthermore, the full counting statistics of the number of particles in a subsystem can be calculated using the expression for  $\det\left[1+\gamma\hat{M}^{\pm}\right]$. The cumulants $\kappa_m$  of the particle distribution function can be found from the cumulants generating function \cite{AIQ11,CMV12}:
\begin{eqnarray}
\kappa_m&=&(-i)^m \left.\frac{\partial^mF(\mu)}{\partial \mu^m}\right|_{\mu=0},\nonumber\\
F(\mu)&=&\ln\det\left[1+(e^{i\mu}-1)\hat{A}\right].
\end{eqnarray}
Using Eq.(\ref{rho_sum}) and Eq.(S9) in the Supplemental Material \cite{suppl}, one arrives at the following result
\begin{eqnarray}\label{cum_gen}
F(\mu)&=&\frac{i\mu}{\pi}(k_FL+\delta^+_{k_F}+\delta^-_{k_F})-\frac{\mu^2}{2\pi^2}\ln(2k_FL)\nonumber\\
&+&2\ln\left(G\left(1+\frac{\mu}{2\pi}\right)G\left(1-\frac{\mu}{2\pi}\right)\right)+o(1),
\end{eqnarray}
where $G(z)$ denotes the Barnes G-function. The average number of particles in the subsystem is then given by
\be
\kappa_1=\frac{k_FL}{\pi}+\frac{1}{\pi}(\delta^+_{k_F}+\delta^-_{k_F}).
\en 
The first term in this expression corresponds to the free fermion case, while the second one describes the change of the average number of particles due to scattering, which is nothing else than the  Friedel sum rule.  Equation (\ref{cum_gen}) shows that all higher order cumulants $\kappa_m$, $m>1$, are not modified by scattering and thus they are the same as in the free fermion case \cite{AIQ11}.

\section{Finite range disordered potential}
\label{disordered}

In order to check that the symmetry of a potential is not essential for our conclusions, we compute numerically  the entanglement entropy of one-dimensional noninteracting  fermions in the presence of a finite range disordered potential. We consider fermions on a lattice:
\be\label{lattice}
-\psi_{n+1}-\psi_{n-1}+(2+V_n)\psi_n=E\psi_n, \quad n=1,\dots,M,
\en 
where $\psi_{0}=\psi_{M+1}=0$ and  $V_n$ are independent random variables uniformly distributed on $[-W/2,W/2]$ for $n\in[(M-a)/2+1,(M+a)/2]$ and  $V_n=0$, otherwise.

For a disorder-free infinite lattice the counterpart of Eq.(\ref{free-fermions}) reads \cite{JK04}
\be\label{free-fermions-lat}
S=\frac{1}{3}\ln(2L|\sin k_F|)+\Upsilon+o(1),
\en
where $k_F$ satisfies $E=2(1-\cos k_F)$. This asymptotic is applicable for a finite lattice provided that $L\ll M$.

\begin{figure}[tb]
\centerline{\includegraphics[width=\linewidth]{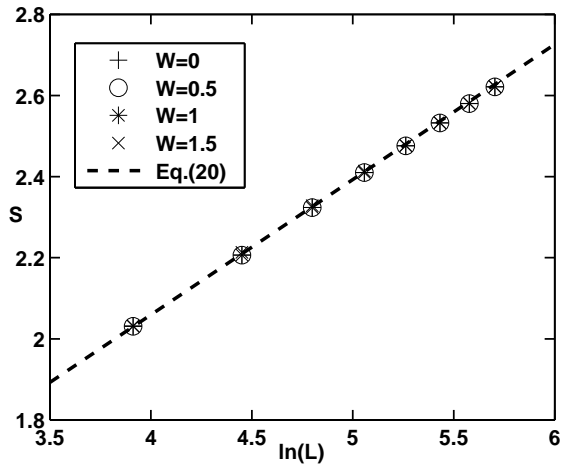}}
\caption{Numerical results for the lattice model (\ref{lattice}) with $M=3000$, $a=10$, $L_{min}=50$ and $L_{max}=300$. Different symbols correspond to different values of the disorder strength $W$, each time $S$ is computed for a single disorder realization. The dashed line is the analytical prediction (\ref{free-fermions-lat}). }
\label{fig_ent}
\end{figure}

Finding the eigenvectors of the Hamiltonian (\ref{lattice}) numerically, we compute $S$ for different values of $L$ and different disorder realizations. The results shown  in Fig.~\ref{fig_ent} clearly demonstrate that the asymptotic behavior of the entanglement entropy is not influenced by a finite range disordered potential, even if the wave functions in the scattering region are localized.

\section{Conclusions}
\label{conclusions}

In this work we have investigated how a scattering potential of finite range $a$ influences the quantum correlations in the ground state of a Fermi gas. The quantity of primary interest is the  entanglement entropy between a finite subsystem of size $L$, which includes the scattering region, and the rest of an infinite system. It can be calculated from the spectral density of the overlap matrix. A general expression for the overlap matrix in terms of the scattering states   is derived first in arbitrary dimensions (\ref{A_general}) and then elaborated in the one-dimensional case (\ref{A-S_relation}). Applied to a symmetric potential, centered at the origin,  it enables us to find analytically the asymptotic behavior of the spectral density in the limit $L/a\to\infty$. Additionally, we compute the entanglement entropy numerically for a generic scattering potential on a lattice.  

Our main conclusion is that the spectral density  of the overlap matrix and hence the von Neumann entropy, the R\'enyi entropies and the full counting statistics are robust against potential scattering in the bulk of a subsystem, provided that $L/a\gg 1$.  Similar universality in the full counting statistics has been reported recently for a Fermi gas confined by a trapping potential \cite{E13}. We note that the scattering at the interface of a subsystem modifies the scaling of the entanglement entropy \cite{P05,ISL09,EP10}. It would be interesting to use our approach to generalize results obtained for that case.

Another interesting variation of the problem considered in this work would be an investigation of the time evolution of the entanglement  entropy after a local quench induced by the scattering potential \cite{CC06}. We expect that in this case the Anderson  orthogonality catastrophe must have a strong impact on the time dependence of the  entanglement  entropy similar to other manifestations of  the Anderson  orthogonality catastrophe in time dependent quenches of Fermi systems such as the Fermi edge singularities.

Finally, we would like to comment on possible experimental verification of our results. The experimental measurement of the entanglement entropy in many-body systems remains an open and very challenging problem. However, there have been several proposals recently suggesting how it can, in principle,  be measured \cite{C11, AD12, DPSZ12}. All of them require preparing  $n$ copies of an original system in order to measure the R\'enyi entropy $S_n$. One of the most promising candidates for such experiments would be a one-dimensional ultracold atomic gas coupled to a two-level atom \cite{AD12}. Remarkably,  the same setup has been proposed recently for an experimental investigation of the Anderson orthogonality catastrophe \cite{GFGPB11, SGGLP13}. Thus, if the experimental realizations of these ideas become feasible, the robustness of the  R\'enyi entropies can be tested.

\acknowledgments

I acknowledge useful discussions with S. Gnutzmann, I. Krasovsky, and I.V. Protopopov, and hospitality of the Abdus Salam ICTP.

\begin{widetext}
\clearpage
\begin{center}
\textbf{\large Supplemental Material: Entanglement entropy in Fermi gases and Anderson's orthogonality catastrophe}
\end{center}
\end{widetext}
\setcounter{equation}{0}
\setcounter{figure}{0}
\setcounter{table}{0}
\setcounter{section}{0}
\setcounter{page}{1}
\makeatletter
\renewcommand{\theequation}{S\arabic{equation}}

\section{Even and odd eigenfunctions and corresponding eigenvalues}\label{app_even_odd}

We consider an integral operator $\hat{A}$ with a kernel $A(k,k')=A(k',k)=A(-k,-k')$ defined on the interval $[-k_F,k_F]$
\be 
(\hat{A}f)(k)=\int\limits_{-k_F}^{k_F}dk'A(k,k')f(k'),\:k\in[-k_F,k_F].
\en
The operator $\hat{A}$ commutes with the parity operator $(\hat{P}f)(k)=f(-k)$ and hence its eigenfunctions are either even or odd functions. We will prove that there is one-to-one correspondence between even (odd) eigenfunctions of $\hat{A}$ and eigenfunctions of $\hat{A}_{+}$ ($\hat{A}_{-}$) defined on the interval  $[0,k_F]$ as 
\be 
(\hat{A}_{\pm}f)(k)=\int\limits_{0}^{k_F}dk'\left(A(k,k')\pm A(k,-k')\right) f(k'),
\en
and the corresponding eigenfunctions have the same eigenvalues.

Let $f_n(k)$ be an  eigenfunction of  $\hat{A}$ with an eigenvalue $\lambda_n$:
\be\label{eigenfunc}
\int\limits_{-k_F}^{k_F}dk'A(k,k')f_n(k')=\lambda_n f_n(k).
\en
Splitting the integral from $-k_F$ to $k_F$ into two integrals from $-k_F$ to $0$  and from $0$ to $k_F$, and changing the integration variable $k'$ by $-k'$ in the first of these integrals, we obtain
\be\label{eigen_also}
\lambda_n f_n(k)=\int\limits_{0}^{k_F}dk'(A(k,k')f_n(k')+A(k,-k')f_n(-k')).
\en
Using that $f_n(-k')=\pm f_n(k')$ for an even (odd) eigenfunction we conclude that an even (odd) eigenfunction of $\hat{A}$ is an eigenfunction of $\hat{A}_{+}$ ($\hat{A}_{-}$) with the same eigenvalue $\lambda_n$.

To prove the converse statement, we take an eigenfunction $f_n(k)$  of $\hat{A}_{+}$ ($\hat{A}_{-}$)
\be
\int\limits_{0}^{k_F}dk'\left(A(k,k')\pm A(k,-k')\right) f_n(k')=\lambda_n f_n(k).
\en
The expression on left hand side can be used to define $ f_n(k)$ for any $k\in[-k_F,k_F]$. The functions constructed in this way are even (odd)
\begin{eqnarray}
&&\lambda_n f_n(-k)=\int\limits_{0}^{k_F}dk'\left(A(-k,k')\pm A(-k,-k')\right) f_n(k')=\nonumber\\
&&\int\limits_{0}^{k_F}dk'\left(A(k,-k')\pm A(k,k')\right) f_n(k')=\pm\lambda_n f_n(k).
\end{eqnarray}
Performing all the transformations that lead from Eq.(\ref{eigenfunc}) to  Eq.(\ref{eigen_also}) in the reverse order, one can see that $f_n(k)$ is an eigenstate of $\hat{A}$ with the same eigenvalue $\lambda_n$.

Now we can apply this statement to the overlap matrix in the case of a symmetric potential.  Its eigenvalues $\lambda_n^{\pm}$ and the corresponding eigenfunctions $f_n^{\pm}(k)$ are solutions of the integral equation
\begin{eqnarray}\label{int_eq_sum}
&&\int\limits_{0}^{k_F}dk'(M^{\pm}(k,k')\pm M^{\pm}(k,-k'))f_n(k')=\lambda_n^{\pm} f_n^{\pm}(k),\nonumber\\
&&M^{\pm}(k,k')=\frac{\sin\left((k-k')L/2+\delta_{k}^{\pm}-\delta_{k'}^{\pm}\right)}{\pi(k-k')},
\end{eqnarray} 
where the property of the phase shifts $\delta_{-k}^{\pm}=-\delta_{k}^{\pm}$ was used \cite{BZP69}. The kernels $M^{\pm}(k,k')$ satisfy the condition 
$M^{\pm}(k,k')=M^{\pm}(k',k)=M^{\pm}(-k,-k')$ and hence, as it follows from the above statement,   the solutions of Eq.(\ref{int_eq_sum}) are in one-to-one correspondence with the eigenvalues and the even or the odd eigenfunctions of the integral operator $\hat{M}^{\pm}$ with the kernel $M^{\pm}(k,k')$ defined on the interval $[-k_F,k_F]$.

The generalized sine kernels $M^{\pm}(k,k')$ can be continuously deformed to the standard sine kernel by replacing the phase shifts $\delta_{k}^{\pm}\to \eta\delta_{k}^{\pm}$ and changing the parameter $\eta$ from $1$ to $0$. It is known that the spectrum of the integral operator corresponding to the standard sine kernel is discrete and the densities of its even and odd eigenvalues are the same \cite{S83}. Assuming that the eigenfunctions change continuously under such deformation, we conclude that the same statement must be correct for $\hat{M}^{\pm}$ and therefore 
\be\label{rho_sum}
\rho(\lambda)=\frac{1}{2}(\rho^{+}(\lambda)+\rho^{-}(\lambda)),
\en
where $\rho^{\pm}(\lambda)$ is the spectral density of  $\hat{M}^{\pm}$.
\section{Spectral density of the generalized sine kernel}\label{dos_gen_sine}
In order to calculate the spectral density of the operators $\hat{M}^{\pm}$, it is convenient to scale the variable $k=k_Fq$ and to consider the operator $\hat{K}$ with the kernel 
\be\label{gen_kernel}
K(q,q',\delta)=\frac{\sin\left((q-q')x+\delta_{k_Fq}-\delta_{k_Fq'}\right)}{\pi(q-q')},
\en
where $q,q'\in[-1,1]$ and $x=k_FL/2$. The spectrum of $\hat{K}$ for $\delta_{k_Fq}=\delta_{k_Fq}^{\pm}$ is the same as for  $\hat{M}^{\pm}$. 

The spectral density of  $\hat{K}$ can be calculated using the corresponding Green's function, which in turn can be extracted from a quotient of two determinants:
\be\label{rho_green} 
\rho(\lambda)=-\frac{1}{\pi}\Im\frac{\partial}{\partial j}\ln\det\left.\left[1-\frac{1}{\lambda-i\epsilon-j}\hat{K}\right]\right|_{j=0},
\en   
where $\epsilon>0$ is an infinitesimal imaginary shift in $\lambda$ and we assume that $0<\lambda<1$. An asymptotic behavior of such determinants was calculated in Ref.\cite{KKMST09}  in the limit $x\to\infty$, provided that $\delta$ is $x$ independent.  Therefore we consider the limit $k_FL\to \infty$ assuming that $k_Fa$ is held constant. 

Applying the main result of Ref.\cite{KKMST09} to the kernel (\ref{gen_kernel}), we obtain
\begin{eqnarray}\label{ln_det}
&&\ln\det\left[1+\gamma\hat{K}\right]=-4i\nu(x+\delta_{k_F})-2\nu^2\ln 4x\nonumber\\
&&+2\ln\left(G(1+\nu)G(1-\nu)\right)+o(1),
\end{eqnarray}
where $\gamma=-{1}/{(\lambda-i\epsilon-j)}$, $\nu=(-1/2\pi i)\ln(1+\gamma)$ and $G(z)$ denotes the Barnes G-function. Then the spectral density can be calculated from this expression following Eq.(\ref{rho_green}). One can notice that the only term, which depends on the scattering phase shift, $-4i\nu\delta_{k_F}$, gives no contribution to $\rho(\lambda)$, as the  imaginary part of its derivative vanishes in the limit $\epsilon\to 0$ \cite{comm_delta}. As a result, the expression for the density of states, that we derive,
\be\label{dos_appendix}
\rho(\lambda)=\frac{\ln 4x - \Re\:\psi\left(\frac{1}{2}+\frac{i}{2\pi}\ln\frac{1-\lambda}{\lambda}\right)}{\pi^2\lambda(1-\lambda)}+o(1)
\en 
coincides with the corresponding  result for free fermions in the absence of scattering \cite{SI12,comment}. The function $\psi(z)$ in the equation above denotes the digamma function.


\begin{thebibliography}{99}
\bibitem{comment} As pointed out in Ref.\cite{SI12} this result is only applicable at $(1/2\pi)|\ln(1/\lambda-1)|\ll k_FL$.
\end{thebibliography}
\begin{thebibliography}{99}
\bibitem{comm_delta} The contribution of this term is $(2/\pi)(\delta(\lambda-1)-\delta(\lambda))$ and it can be discarded for $0<\lambda<1$.
\end{thebibliography}
\end{document}